\documentstyle[aps,multicol,epsfig,eqsecnum]{revtex}
  
\begin{document}
\draft

\title{Evolution of polymorphism and sympatric speciation through competition 
in a unimodal distribution of resources}

\author{E. Brigatti $^{\dag\pm}$,
         J.S. S\'a Martins $^{\star}$   
         and I. Roditi $^{\dag}$ 
        }  
  
\address{$\dag$Centro Brasileiro de Pesquisas F\'{\i}sicas, Rua Dr. Xavier 
  Sigaud 150, 22290-180, Rio de Janeiro, RJ, Brasil} 
\address{$\star$Instituto de F\'{\i}sica, Universidade Federal Fluminense, 
  Campus da Praia Vermelha, 24210-340, Niter\'oi, RJ, Brasil}
\address{$\pm$e-mail address: edgardo@cbpf.br}

\maketitle

\widetext
  
\begin{abstract}
A microscopic agent dynamical model for diploid age-structured populations 
is used to study evolution of polymorphism and sympatric speciation. The 
underlying ecology is represented by a unimodal distribution of resources of 
some width.  Competition among individuals is also described by a similar 
distribution, and its strength is maximum for individuals with the same 
phenotype and decreases 
with distance in phenotype space as a gaussian, with some width. These two 
widths define the model's phase space, in which we identify the regions where 
an autonomous emergence of stable polymorphism or speciation is more likely.

\end{abstract}
  
\pacs{87.23.Kg, 87.23.-n, 05.10.Ln}
   


\begin{multicols}{2}

\section{Introduction}

The genesis of the incredible diversity of life on our planet is the problem 
that lies at the heart of the development of modern theory of evolution. This 
theory attempts to describe all its diverse processes, and in particular 
speciation, as the outcome of a microscopic dynamics driven by selection and 
mutation. Insofar as the biological definition of species goes, speciation is 
the process that explains the generation of two reproductively isolated 
populations, for which gene flow between the different taxa is absent in any 
form. The most intuitive conjecture as to how speciation can occur leads to a 
scenario in which a geographical barrier, or any physical isolation 
mechanism, separates some fraction of the population of a species. This 
process is called allopatric speciation. This description is nowadays well 
accepted and supported by an abundance of empirical evidence \cite{alop,Coyne}. 
Much more subtle and complex is the conjecture that speciation is also 
possible in the absence of any physical isolation mechanism: a single 
continuous population of interbreeding organisms can split into two 
reproductively isolated subpopulations. This conjecture is not new, dating at 
least from Darwin himself \cite{Charles}, albeit from a perspective that is 
somewhat different from the modern one. A great interest in this theme has 
grown lately \cite{Maynard,Lande}, strongly supported by observational data related 
to some fishes of small crater lakes of Cameroon \cite{Schliewen} and by 
recent papers that demonstrate sympatric speciation in vitro \cite{Friesen} 
or in artificial-life type simulations \cite{Chow}. On the other hand, the 
effort towards the construction of general models that can give structure and 
plausibility to verbal theories (see Ref. \cite{Coyne,Turelli} for a general 
review) have caused the investigation of an extensive number of mathematical 
and computational models. Out of the wide variety of such models, we restrict 
our analysis to ecology-driven reproductive isolation, and we address our 
attention to models in which natural selection depending on resources 
distribution originates sympatric speciation. Some of the early attempts on 
these lines were made with models that represent heterogeneous environments 
with two niches, and where a mechanism of adaptation to such discrete 
resources was established. In this case, speciation is caused by a fixed 
selection that favors phenotypes of both the extremes of the possible range 
(disruptive selection) and leads to reproductive isolations. A model with 
these ingredients appeared recently in the literature \cite{KK}, and a 
similar one, based on microscopic dynamics, was studied by one of the authors 
\cite{Jorge,Karen}. The reasoning that underlies these models suggests the 
question of whether discrete niche might be necessary to explain species 
coexistence in sympatry. A recent analysis \cite{Dieckmann} tried to give a 
negative answer to this question, showing how speciation could arise from 
competition for continuously distributed resources. In such situations, the 
environment has just a single ecological niche and, as a consequence, the 
population feels a stabilizing selection that favors intermediate phenotype. 
In this scenario, the introduction of a competition mechanism generates a 
disruptive selection that weakens as species diverge. This competition is 
generated through a selection force that is stronger for more frequent 
phenotypes and weaker for rare ones (frequency-dependent selection). 
This mechanism is interesting for two different reasons. First, such a sequence 
of events is controlled by a 
frequency-dependent interaction and does not require any externally imposed 
disruptive selection pressure. On top of that, it appears to be a more 
general mechanism, which can occur under a much wider range of ecological 
situations. These two selection pressures are yet not enough to generate 
speciation. To obtain a stable process, it is necessary to prevent the 
appearance of intermediate phenotypes through the divergence of reproductive 
compatibility. This is obtained thanks to the evolution of assortativity,
a non-random mating strategy. 
In the case of a unimodal distribution of resources, such a mechanism is also 
strictly necessary for the appearance of a first phenotypic 
divergence. All these ideas have been recently developed in Ref. 
\cite{Dieckmann} and are the point of departure for our present analysis, 
where we want to study all these results for a more complex microscopic model.
 
 \section{Model and methods}
 
\begin{figure}[p]
\begin{center}
\vspace*{0.8cm}
\resizebox{0.4\textwidth}{!}{\includegraphics{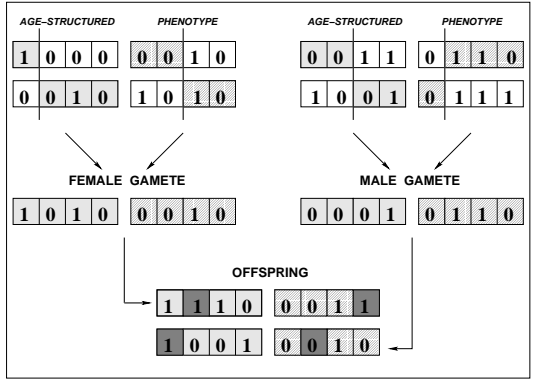}} 
\vspace*{0.4cm}
\end{center}
\caption{\small Example of a reproductive cycle. 
The diploid genome is represented
with its age-structured part (light-shaded background) 
and the bit-strings that encode 
for phenotype (diagonal stripes background).   
After performing crossing-over in the first passage,
the haploid gamete is chosen in the second.
Finally, some new mutation are added 
(dark-shaded squares) and the gametes 
combine to form a new individual.}
\label{fig_Exa}
\end{figure}

We are interested in working at a microscopic level, where the fundamental 
agent of the model is an individual of an age-structured population with 
phenotype variability. For this reason, the classical Penna model 
\cite{Penna} is a convenient point of departure, in its asexual haploid and 
sexual diploid versions. In the following, we will describe results relative 
to both sexual and asexual taxa. In the first case, we are dealing with the 
onset of sympatric speciation in a strict sense, which coincides with the 
setting up of reproductive isolation. In the second, we are investigating the 
emergence of polymorphism, meaning the splitting of a phenotypical 
monomorphic population into two (or more) distinct phenotypic clusters. The 
sexual case describes a diploid population, representing each individual 
through two sets of coupled genetic strands. Each one is built up by two 
$32$-bit long bit-strings. The first string is age-structured and is used to 
introduce the biological clock of the individuals, while the other represents 
the phenotype (trait bit-string). In each of these strings, a gene is encoded 
by two homologous bits. Its value is $0$ if it describes an allele equal to 
the one present in the original population - wild type allele - or $1$ if it 
has suffered a change and has become a mutant allele. Each individual of a 
reproductively active couple generates, through a meiotic cycle with 
crossing-over and recombination, one haploid gamete that, after the 
introduction of some mutations, combine to form a new genetic strand  (see Fig. 
\ref{fig_Exa}). 

The phenotype value is associated with the second string of the genome by the 
overall sum of the active mutations present in that portion. By active 
mutation we mean a homozygous locus, where the bit value of the two 
homologous alleles are equal to one, or a heterozygous one for which the 
mutant allele is dominant. The number of loci where the  mutant alleles are 
dominant is fixed and their position is chosen randomly at the beginning of the 
simulation. According to this procedure, the phenotype value 
($x$) is an integer value between $0$ and $32$, and is different from the 
simple sum of all the ones present in the bit-string. This kind of 
characterization of the phenotype takes into account the diploid nature of 
the genome and is in some way consistent with the representation of a 
quantitative trait. This must be understood in the sense that we are 
representing a trait depending on multiple genes (polygenic trait), but in a 
particularly simplified situation: the environment has a negligible effect on 
the phenotype and the quantitative trait is ideal (the genes are randomly 
associated and their effect is completely additive \cite{Daniel}). 

The genetic and, in correspondence, phenotypic variability is assured by the 
action of mutations. Mutations are obtained by randomly flipping one of the 
bits of the string. Bits of the trait bit-strings can mutate from $0$ to $1$ 
or from $1$ to $0$, as opposed to the ageing bit-string that can undergo only 
bad mutations. This is an oversimplification: for the trait bit-string we do 
not take into account that it is really difficult for a mutated allele to 
restore its previous activity. On the contrary, for the ageing bit-string we 
consider only harmful mutations because of their strong predominance in 
nature. With the use of this phenotype characterization it is possible to 
establish a computational representation of intraspecific and/or 
environmental interaction and/or sexual selection  \cite{Jorge,Karen,Edgardo}. 

To obtain an age structure in the population, we allow each agent to live 
until the occurrence of death caused by ageing. This is implemented following 
rules inspired by Medawar's hypothesis of the accumulation of bad mutations. A 
position (locus) of the chronological (age-structured) piece of the genome is 
read at each time step. If an active mutation (defined in the same manner as 
for the phenotype value) is found at this locus, it is added to the current 
number of harmful mutations; the individual dies when this amount reaches 
some pre-determined threshold value. 

In the model we introduce another death factor that represents the natural 
causes of selection in a real ecosystem. It is characterized by three 
different components. The first is density-dependent, responsible for 
limiting the number of the total population in accordance with a logistic 
growth. Then, a frequency-dependent factor takes into account how, in 
realistic situations, the tendency to occupy the more favored regions in 
phenotype space is contrasted to an increasing competition between 
individuals. To these dynamic components of selection, which represent the 
feedback between individuals and ecosystem and take into account the 
evolution of the environment, a static component is added. It designs the 
general ambient condition of the territory and the ecological niche where the 
agents live. This is a cause of a directional selection that constantly 
drives the population towards a fitness maximum in phenotype space. To sum 
up, the general expression for such death factor, as used 
in the Monte Carlo simulations, is: 
\begin{eqnarray}
V = \frac{\sum_{y=1}^{32} N_{y}\cdot\exp(-(x-y)^2/2c^2)}
{k \cdot \exp(-(x-16)^2/2s^2)}\nonumber
\end{eqnarray} 
At each time step in the simulation a random number is tossed; the individual 
survives if this number is larger than the $V$ value. The static fitness 
landscape is represented by a gaussian with deviation $s$ and drives all the 
population towards the fitness maximum equal to $16$. The competition 
declines with phenotype distance according to a gaussian function with 
deviation $c$, and the parameter $k$ controls the population dimension. $x$ is 
the phenotype value of the individual that is feeling the selection pressure 
and the sum runs over the $y$ index that spans all of the phenotype space. 
With $N_{y}$ we indicate the number of individuals with phenotype $y$.

As already mentioned, positive assortative mating plays a basic role in the path 
towards speciation. This is 
obtained by allowing mating only if the distance in phenotype space between 
two partners is smaller than some predefined threshold $a$ \cite{Franco}. 
It is interesting to notice that several models (for example in \cite{Dieckmann}) 
implement some mechanism that, by normalizing mating probabilities, guarantees 
an equal reproductive success for all individuals. This care is due to the fact 
that assortative mating can cause a sexual selection on rare phenotypes, that have 
a smaller possibility of finding mates. This byproduct of assortative mating, 
which has a stabilizing effect that makes speciation less likely 
\cite{Kirkpatrick}, can be present in nature. For this reason, in our model, we do 
not implement any mechanism for equalizing mating success. 

Finally, we are not focusing on the problem of the natural evolution of 
assortativity 
\cite{Dieckmann}; in contrast, we are assuming here a fixed assortative rule 
from the start, for the sake of simplicity. In fact, with such a schematic 
mechanism, the tuning of the parameter that leads to reproductive isolation 
is a hard problem.

\section{Results}

We begin the presentation of our results by showing some general data for an 
asexual population. The dynamical rules for such a population are exactly the 
same as for the sexual version, except for the fact that we are now dealing 
with a haploid population, with a very simple reproductive cycle, and where 
the mutation process stands as the only cause of variability \cite{Penna}.
   
\begin{figure}[p]
\begin{center}
\vspace*{0.8cm}
\resizebox{0.4\textwidth}{!}{\includegraphics{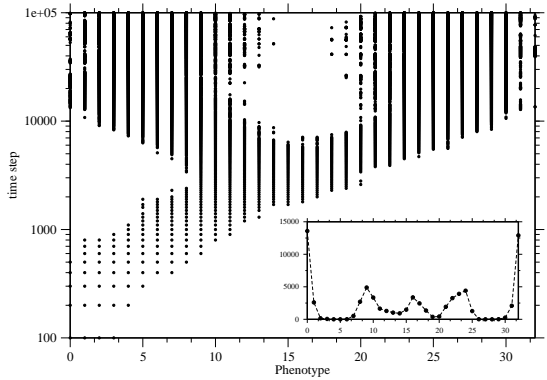}} 
\vspace*{0.4cm}
\end{center}
\caption{\small Time evolution of an asexual population. The parameters used 
in the simulations are: $k$ ($100,000$), the initial population ($1,000$), 
the minimum reproduction age ($6$), the maximum reproduction age ($32$), the 
number of offspring per mating season ($2$), the threshold value for harmful 
diseases ($3$), the number of mutations added at birth in the age structured 
string ($1$), and the mutation probability in the trait string ($0.01$). The 
values for the standard deviation of the competition ($c$) and of the static 
component ($s$) are respectively $20$ and $25$. The inset shows the unstable 
distribution generated by a simulation with $s=25$ and $c=5$ after $100,000$ 
time steps.}
\label{fig_Dyn}
\end{figure}

\begin{figure}[p]
\begin{center}
\vspace*{0.8cm}
\resizebox{0.4\textwidth}{!}{\includegraphics{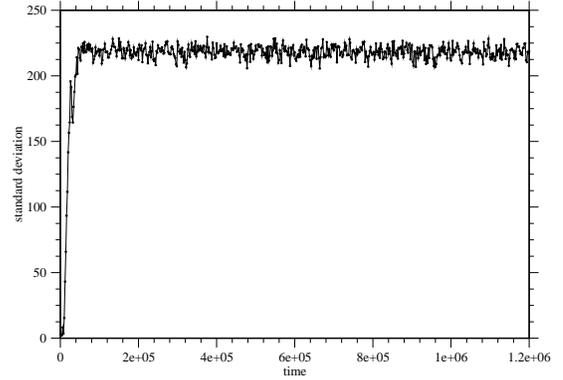}} 
\vspace*{0.4cm}
\end{center}
\caption{\small The standard deviation of the phenotypes distribution can be 
considered as an order parameter: the splitting of a phenotypical monomorphic 
population into two (or more) distinct phenotypic clusters corresponds to an 
abrupt increase in the value of the standard deviation. The fact that after 
the first transition, near $t=30000$, the mean value of the standard 
deviation does not change anymore proves the stability of the bimodal 
phenotypic distribution ($s=25$, $c=20$).}
\label{fig_Long}
\end{figure}

\begin{figure}[p]
\begin{center}
\vspace*{0.8cm}
\resizebox{0.4\textwidth}{!}{\includegraphics{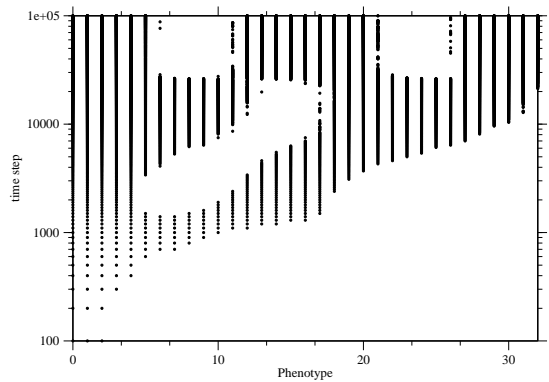}}
\vspace*{0.4cm}
\end{center}
\caption{\small A particular situation where it is possible to rapidly obtain 
two speciation events ($s=40, c=6, a=4$). }
\label{fig_Dyn3}
\end{figure}

In this simple case the population evolves rapidly towards the fittest 
phenotype ($16$) that corresponds to the maximal carrying capacity. Once this 
value has been reached, two scenarios are possible. In the first, $16$ is a 
stable point in phenotype space and the population will structure itself in a 
well shaped gaussian distribution around such value. Otherwise, $16$ is still 
not a fitness maximum (see figure \ref{fig_Dyn}), and the population 
experiences disruptive selection that leads to the appearance of two distinct 
phenotypic morphs. In this latter case, the tendency to avoid an overcrowded 
region of phenotype space is stronger than the advantage obtained by reaching 
the optimal phenotype value. This happens when $c<s$ (see table 
\ref{tab_Asex}). This is not the only possible behavior though; for some 
small enough values of $c$, the population rapidly splits up its phenotype 
distribution before its mean reaches the fittest phenotype. An example of 
this situation is given in the sexual case (see Fig. \ref{fig_Dyn3}). This 
particular behavior is related to a stronger influence of competition, that 
drives a premature bifurcation. In fact, this force increases as $c$ is 
decreased, as also reflected in Table \ref{tab_Asex} by a larger number of 
occurrences of branching events. This is easily seen, for instance, when 
$c = 10$ where, in just $100000$ time steps the population undergoes two 
bifurcation events, each one causing the appearance of two distinct phenotype 
clusters, leading towards a three-modal distribution. As we increase the 
competition further, the population feels such a strong drive that it is 
impossible to reach a stable polymorphism: the branching events are now so 
numerous that the phenotype distribution becomes unstable, characterized by a 
large number of peaks connected by intermediate phenotypes (see inset in 
figure \ref{fig_Dyn}). These results are comparable with the ones obtained in 
Ref. \cite{Dieckmann}, but the assignment of the phenotype from a genome ruled by 
microscopic dynamics determines a richer phenomenology. 
In fact, their prediction of branching for 
$c<s$ is confirmed by our study, but we were also able to show how branching 
can occur repeatedly, leading to a polymorphism with more than two phenotype 
clusters, or to even more complex unstable behavior, as reflected in the 
enriched unfolding of our model in time. 
Moreover, data obtained by longer simulation show how the clustering of the 
population phenotype is biological relevant. In fact, also after $10^6$ time steps 
polymorphism is still present (see figure \ref{fig_Long}). For this reason, even if from a numerical 
simulation it is impossible to determine the final equilibrium distribution, such a 
long-lived structure can not be considered a simple transient phase 
\cite{Polechova} but, at least, as some sort of long-standing quasi-stationary 
state with a time scale comparable to the ones of biological interest. 

\begin{figure}[p]
\begin{center}
\vspace*{0.8cm}
\resizebox{0.4\textwidth}{!}{\includegraphics{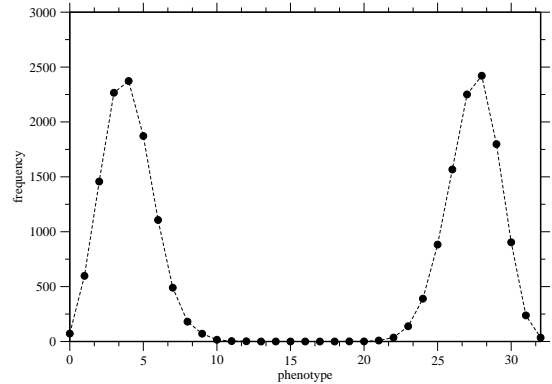}} 
\vspace*{0.4cm}
\end{center}
\caption{ \small An example of speciation in a sexual population. The figure 
shows the phenotype distribution after $100,000$ time steps. The parameters 
used in the simulations that are different from those of Figure \ref{fig_Dyn} are: 
$s=40$, $c=8$, $a=4$, the minimum reproduction age ($8$), the number of 
offspring per mating season ($4$), and the number of loci where the $1$ allele 
is dominant in each string ($16$).}
\label{fig_branch}
\end{figure}

Evolution of diploid sexual populations is marked by a stronger variability, 
due to the effects of Mendelian segregation and recombination. This fact is a 
source of difficulties in obtaining sharp phenotypic differentiations because 
of the constant generation of intermediate phenotypes. For this reason, if the 
mating process is random, a stable splitting into two different phenotypic 
clusters can no longer occur, for any possible value of $s$ and $c$. On the 
contrary, with the introduction of positive assortative mating, not only it is 
possible to recognize different phenotypic morphs but also speciation with 
sexually isolated populations (Fig. \ref{fig_branch}). The parameter space 
where it occurs appears to be more confined than in the asexual cases. 
Although still necessary, it is no longer sufficient that $c<s$. Out of the 
various simulations performed, we have been able to obtain speciation for 
some simulations with parameters set in the following region: 

\begin{eqnarray}
s &\ge& 15,  6 \le c \le 8, a=4;\nonumber \\
s&=&40, c=6, 3 \le a \le 5 \nonumber
\end{eqnarray}

\begin{figure}[p]
\begin{center}
\vspace*{0.8cm}
\resizebox{0.4\textwidth}{!}{\includegraphics{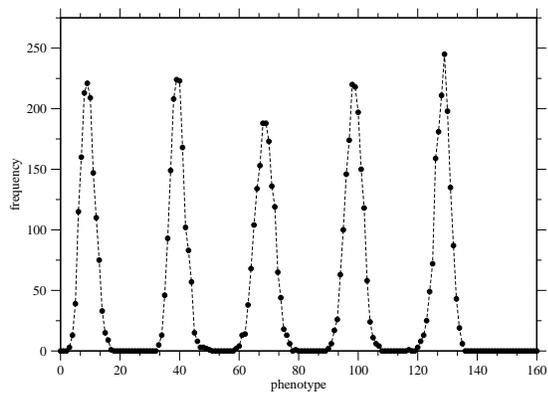}} 
\vspace*{0.4cm}
\end{center}
\caption{ \small A simulation with a trait 
bit-string of $160$ loci ($s=160$, $c=8$, $a=4$).  
This result suggests that there is no dependence between speciation and trait 
range, in spite of a recent claim to the contrary [19]. } 
\label{fig_branch160}
\end{figure}

\noindent
The simulations show that the system is not very sensitive with respect to 
the value of $s$. If it is sufficiently larger than a particular value of 
$c$, the population is able to escape from the region with maximum carrying 
capacity. On the other hand, the value of $c$ can neither be too small, which 
would lead to a distribution characterized by not well defined connected 
peaks, nor too large, otherwise the competition would not be strong enough to 
drive the distribution away from a gaussian centered in $16$. Finally, values 
of $a$ that are too small can totally trap the variability of the phenotype 
distribution to a confined region close to $0$ or even prevent speciation because 
of the stronger penalizing effect of sexual selection on rare phenotypes 
\cite{Kirkpatrick}. 
For larger values, on the 
opposite, the mating choice is not strict enough to prevent intermediate 
phenotypes. In any case, even in this more restricted scenario, it is still 
possible to find some more favorable situations for which, after just 
$100,000$ time steps, two speciation events have already occurred (see figure 
\ref{fig_Dyn3}). With respect to the dependence on the initial condition, we 
remark that if we start with a population having a phenotype close to $16$ it 
is, in general, easier to obtain speciation. We again claim our results to be 
in accordance with those of Dieckmann and Doebeli \cite{Dieckmann}, but the 
parameter requirement for the emergency of speciation appears to be more 
restrictive, even with the usage of a stronger and fixed mating rule. We have 
also performed some simulations with a model characterized by a big trait 
bit-string (up to $160$ loci) \cite{Polechova}. The fact that no qualitative 
differences emerged suggests that the results obtained are not dependent on the 
phenotype range (see Fig. \ref{fig_branch160}). 

\section{Conclusions}

The representation of an individual phenotype character through a 
second bit-string is a natural extension of the classical Penna model that allows 
the study  of sympatric speciation. Our model was structured under these 
guidelines and the results corroborate the hypothesis that sympatric speciation 
emerges from competition for continuously distributed resources. We also used it 
to test the robustness of this hypothesis under the inclusion of further 
ingredients and the dependence of the results on the values of some of the model's 
constitutive variables.
  
With the study of an age-structured population we can substantiate the claim that 
the precedent results were not biased by possible long-lived individuals.
Moreover, in the asexual population the introduction of a phenotype determined by 
a genotype with $32$ loci confirm previous results \cite{Dieckmann} in an enriched 
scenario. The fact that clustering is still present, even after more than $10^6$ 
time steps, determines the biological relevance of such a long-lived phase.
For a sexual population where the mechanisms of crossing-over, dominance and 
homo/heterozygous loci were implemented, the emergence of sympatric speciation is 
obtained. Our results, obtained with a model that does not include any mechanism 
for equalizing mating success, contradict the alleged stabilizing effect due to 
sexual selection caused by assortative mating. Finally, the agility of the Penna 
model in manipulating bit-strings allowed the run of simulations with large 
phenotype range, with results that suggest that there is no dependence between 
speciation and trait range, in spite of a recent claim to the contrary 
\cite{Polechova}. It is our opinion that agent-based models that allow statistical 
fluctuations, such as the one in Ref. \cite{Dieckmann} and ours, are the most 
promising testing grounds for evolutionary ideas as a whole.

\section*{Acknowledgments} 

We thank the Brazilian agencies CAPES, CNPq, and grants from PRONEX 
(PRONEX-CNPq-FAPERJ/171.168-2003 and PRONEX-FAPERJ E-26/171/2003) and FAPERJ 
(E-26/170.699/2004) for partial financial support.

\newpage

\begin{table}[p]
\begin{center}
\begin{tabular}{cc|ccccccc}
               & 40 & 1 & 1 & 1 & 1 & 1 & 1 & 1\\
     {\bf $c$} & 35 & 1 & 1 & 1 & 1 & 1 & 1 & 2\\
               & 30 & 1 & 1 & 1 & 1 & 1 & 2 & 2\\
               & 25 & 1 & 1 & 1 & 1 & 2 & 2 & 2\\
               & 20 & 1 & 1 & 1 & 2 & 2 & 2 & 2\\
               & 15 & 1 & 1 & 2 & 3 & 3 & 3 & 2\\
               & 10 & 1 & 3 & 3 & 3 & 3 & 3 & 3\\
               & 5  & x & x & x & x & x & x & x\\
\hline
           & & 10 & 15 & 20 & 25 & 30 & 35 & 40\\
                & &  &  &  &  &  & {\bf $s$} & \\                           
\end{tabular}
\end{center}
\caption{ \small Generation of polymorphism in an asexual population as we vary 
the values of the standard deviation of the competition ($c$) and of the static 
component ($s$). Each number in the table corresponds to the number of isolated 
peaks in the phenotype distribution, while a cross indicates the occurrence of 
unstable polymorphism, such as the one represented in the inset in figure 
\ref{fig_Dyn}. These results are the outcome of one simulation. 
We have performed a number of different simulations varying the initial condition. 
In all cases, the appearance of distinct phenotypic morphs happens only when 
$c<s$, and the overall dependence of the results on the parameters is similar, 
even when the number of peaks, isolated or not, vary quite a lot. 
The simulations run with the same parameters as those of figure \ref{fig_Dyn}.
The picture was taken after $100,000$ time steps, after all 
distributions have reached a stationary situation.}
\label{tab_Asex}
\end{table}

\end{multicols}

\end{document}